\documentclass[twocolumn]{jpsj2} 
\newcommand{\Romannum}[1]{\uppercase\expandafter{\romannumeral#1}}
\usepackage{bm}
\usepackage{mathrsfs}
\usepackage{exscale}
\usepackage{color}
\title{Two-Dimensional Spin Dynamics in the Itinerant Ferromagnet LaCoPO \\
Revealed by Magnetization and $^{31}$P-NMR Measurements}
\author{
Hitoshi \textsc{Sugawara}$^{1}$, 
Kenji \textsc{Ishida}$^{1,2}$\thanks{E-mail: kishida@scphys.kyoto-u.ac.jp}, 
Yusuke \textsc{Nakai}$^{1,2}$\thanks{E-mail: nakai@scphys.kyoto-u.ac.jp},\\
Hiroshi \textsc{Yanagi}$^{3}$,
Toshio \textsc{Kamiya}$^{3,4}$,
Yoichi \textsc{Kamihara}$^{2}$,
Masahiro \textsc{Hirano}$^{4,5}$,
and 
Hideo \textsc{Hosono}$^{3,4,5}$
}

\inst{$^{1}$Department of Physics, Graduate School of Science, Kyoto University, Kyoto 606-8502, Japan\\
$^{2}$TRIP, JST Sanban-cho Building, 5 Sanban-cho, Chiyodaku, Tokyo 102-0075, Japan\\
$^{3}$Materials and Structures Laboratory, Tokyo Institute of Technology, Yokohama 226-8503, Japan\\
$^{4}$ERATO-SORST, Japan Science and Technology Agency (JST), Yokohama 226-8503, Japan\\
$^{5}$Frontier Research Center, Tokyo Institute of Technology, Yokohama 226-8503, Japan\\}

\abst{
We have performed magnetization and $^{31}$P-NMR measurements on the itinerant ferromagnet LaCoPO (Curie temperature $T_{\rm Curie}\sim 44$ K) with a layered structure in order to investigate spin dynamics in the paramagnetic state. The linear scaling between the Knight shift $K$ at the P site and the bulk susceptibility $\chi$ above $T_{\rm Curie}$ indicates that the P nucleus is suitable for investigating magnetic properties. The temperature and magnetic field dependences of the nuclear spin-lattice relaxation rate divided by the temperature $1/T_1T$ at the P site show characteristic features of itinerant ferromagnets, such as ZrZn$_2$ and Y(Co$_{1-x}$Al$_x$)$_2$. In addition, the relationship between $1/T_1T$ and $\chi$ above $T_{\rm Curie}$ suggests that ferromagnetic fluctuations possess a two-dimensional (2D) characteristic. The present data show that LaCoPO is a unique ferromagnet, where the 2D fluctuations anticipated from the crystal structure are predominant down to almost $T_{\rm Curie}$.  
}

\kword{Cobalt pnictide, ferromagnetism, two dimensional, ferromagnetic fluctuations}

\begin{document}
\maketitle

Since the discovery of superconductivity in LaFeAs(O$_{1-x}$F$_x$) with the ``1111'' structure [tetragonal ZrCuSiAs structure ($P4nmm$)]\cite{KamiharaFeAs}, much attention has been paid to the electronic state in the two-dimensional (2D) FeAs layer where superconductivity occurs. In the FeAs compounds with the ``1111'' structure, the presence of cylindrical Fermi surfaces with hole and electron characteristics was inferred from the band calculations\cite{Lebegue,SinghPRL2008} and has been confirmed by angle-resolved photoemission spectroscopy (ARPES) measurements in NdFeAs(O$_{1-x}$F$_x$)\cite{LiuNdFeAsOF}. The antiferromagnetic (AFM) order at $T_N \sim 140$ K in LaFeAsO is considered to be induced by nesting between the hole and electron Fermi surfaces\cite{CruzNature2008}, which is believed to be related to high-$T_{\rm SC}$ superconductivity in 2D FeAs compounds\cite{MazinPRL2008,KurokiPRL2008}. Magnetism in the ``1111'' structure other than that in the FeAs compounds is also important for exploring a possible candidate for a new superconductor.

In this paper, we report magnetic properties in LaCoPO with the ``1111'' structure. LaCoPO shows a ferromagnetic (FM) transition at 44 K with a small spontaneous magnetic moment $p_s = 0.33~\mu_B/$ Co\cite{YanagiPRB08}. The resistivity shows a metallic behavior from room temperature down to 2 K and the effective magnetic moment estimated from the Curie-Weiss behavior of the susceptibility $\chi$ just above $T_{\rm Curie}$ is approximately 2.9 $\mu_B$, and thus, LaCoPO is regarded as a weak itinerant ferromagnet. Since related compounds LaFePO and LaNiPO with even numbers of $3d$ electrons [Fe(3$d^6$) and Ni(3$d^8$)] show superconductivity\cite{KamiharaFeP,Tegel2008193,LaNiOWatanabe} and LaMnPO with Mn(3$d^5$) is an antiferromagnet\cite{LaMnPO}, ``1111'' compounds with odd-number 3$d$ electrons may exhibit a magnetic ground state. Measurements of the bulk susceptibility and $^{31}$P-NMR ($I$ = 1/2) have been performed on LaCoPO in order to investigate magnetic properties and spin dynamics in the paramagnetic (PM) state. The linear scale between the bulk susceptibility and the Knight shift ($K$) at the P site indicates that $K$ is determined by Co-$3d$ spins, and that a $^{31}$P nucleus is suitable for probing Co spin dynamics. By the analyses of the nuclear spin-lattice relaxation rate $1/T_1$ and $\chi$ above $T_{\rm Curie}$, we obtained the relation $1/T_1T \propto \chi^{3/2}$ in the PM state, suggestive of the predominance of the 2D FM fluctuations in LaCoPO.

The polycrystalline powder LaCoPO was employed in our measurements. Detailed preparation and characterization are described in ref.\cite{YanagiPRB08}. The magnetization was measured with a commercial superconducting quantum interference device magnetometer (Quantum Design, MPMS-XL).
\begin{figure}[htbp]
\begin{center}
\includegraphics[clip=,width=0.8\columnwidth]{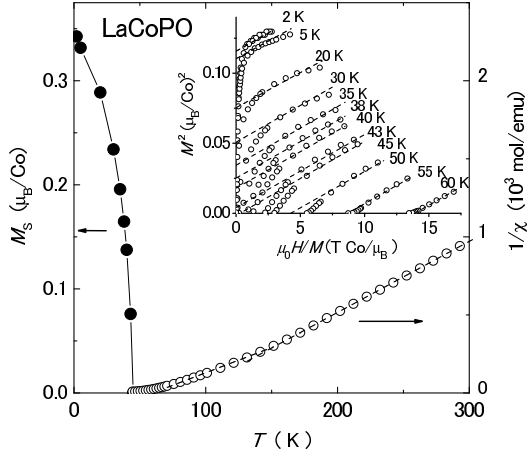}
\end{center}
\caption{Temperature dependence of spontaneous magnetization $M_{\rm S}(T)$ ($\mu_B/$Co) and reciprocal magnetic susceptibility $1/\chi$ (mol/emu) in LaCoPO. $M_{\rm S}(T)$ was determined with the Arrott plot shown in the inset, and the $1/\chi$ in the paramagnetic state was determined from $\lim_{H \rightarrow 0} H/M$. The solid curve in $M_{\rm S}(T)$ is a guide for the eye, and the broken lines in $1/\chi$ are the Curie-Weiss fits for the evaluation of $P_{\rm eff}$ (see text).The inset shows the plot of $M^2$ versus $H/M$ (Arrott plot) at various temperatures. The dotted lines in the inset represent linear relations for evaluating $T_{\rm Curie}$ and $M_S(T)$.}
\end{figure}
Figure 1 shows the temperature dependence of the spontaneous magnetization $M_{\rm S}(T)$ and $1/\chi(T)$ determined through the Arrott plots shown in the inset, where the square of the isothermal magnetization $M^2$ is plotted against the inverse of the magnetic susceptibility ($H/M$). A deviation from a linear relation is observed at small $H/M$ below 50 K. This convex behavior was also reported in LaCoAsO\cite{OhtaPRB09}, which is considered to originate from the sum rule of the zero-point and thermal spin fluctuations\cite{TakahashiJPSJ86} and/or the two-dimensional characteristic of the spin fluctuations. The Curie temperature $T_{\rm Curie}$ and spontaneous magnetization $M_{\rm S}$ shown in Fig.~1 were determined by the extrapolation of the linear relation and evaluated to be $T_{\rm Curie} = 44 \pm 1$ K and $M_{\rm S} = 0.33 \pm 0.02 \mu_B$/Co at 2 K, respectively. These values are in good agreement with those in our previous report\cite{YanagiPRB08}. $1/\chi(T)$ above $T_{\rm Curie}$ is derived from $\lim_{H\rightarrow0}H/M$. The effective Curie moments $P_{\rm eff}$ are $1.57 \pm 0.02~\mu_B$, as obtained by the fitting of $1/\chi(T)$ in the temperature range between 80 and 150 K, and $1.30 \pm 0.01~\mu_B$ between 150 and 300 K. The obtained $P_{\rm eff}$ is smaller than that previously reported of LaCoPO (2.9 $\mu_B$), but it is very similar to that of LaCoAsO ($1.3 \sim 1.4 \mu_B$), smaller than the theoretically expected value of 3.87 $\mu_B$ for $S = 3/2$. At present, the cause of the large value of $P_{\rm eff}$ reported previously is unclear, but it seems to be a spurious impurity phase. 

\begin{figure}[htbp]
\begin{center}
\includegraphics[clip=,width=0.8\columnwidth]{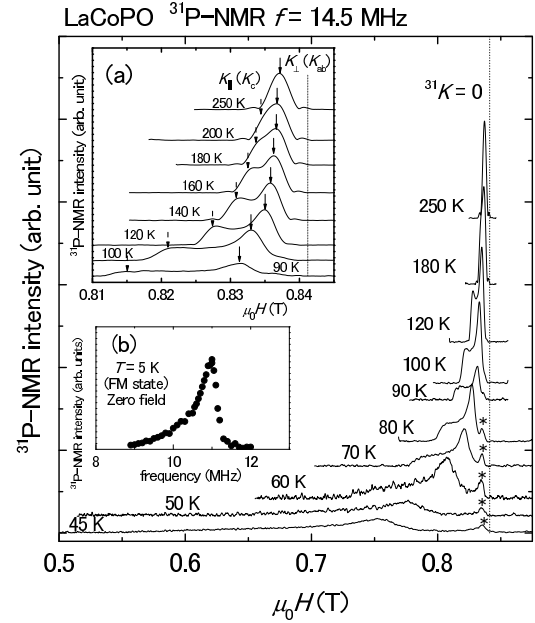}
\end{center}
\caption{Field-swept $^{31}$P-NMR spectra obtained at 14.5 MHz in a series of various temperatures above $T_{\rm Curie}$. The small peaks marked by asterisks are assigned to an impurity phase since the peaks are temperature-independent. $K = 0$ is determined with respect to the $^{31}$P resonance line in H$_3$PO$_4$. The inset (a) shows the close view of the NMR spectra at high temperatures. The inset (b) shows the zero-field $^{31}$P-NMR spectrum observed at 5 K far below $T_{\rm Curie}$.}
\end{figure}
$^{31}$P-NMR ($I$=1/2) measurements were performed at 4.8, 14.5, and 28.5 MHz, in which $^{31}$P-NMR spectra were observed at around 0.28, 0.84 and 1.70 T, respectively. Figure 2 shows the P-NMR spectra at 14.5 MHz obtained by sweeping magnetic fields at various temperatures. To observe their variations carefully, the inset (a) shows the extended $^{31}$P-NMR spectra at high temperatures. The P-NMR spectrum at $T$ = 160 K shows a typical powder pattern spectrum with the axial Knight shift component. The Knight shifts perpendicular $K_{\perp}$ and parallel $K_{\parallel}$ to the axis of the axial symmetry, which is the $c$-axis in LaCoPO, are determined from the resonance spectrum, as shown by the solid and dotted arrows, respectively. $K_{c}$ shifts markedly with decreasing temperature, and the spectra become anomalously broader toward $T_{\rm Curie}$. At 5 K well below $T_{\rm Curie}$, a zero-field NMR signal was observed at around 11 MHz, as shown in the inset (b), where the resonance magnetic field purely arises from the FM Co moments. From this resonance frequency, the magnitude of the ordered moments at the Co sites can be roughly estimated using the hyperfine coupling constant between $^{31}$P nuclear and Co electronic spins.   

\begin{figure}[htbp]
\begin{center}
\includegraphics[clip=,width=0.8\columnwidth]{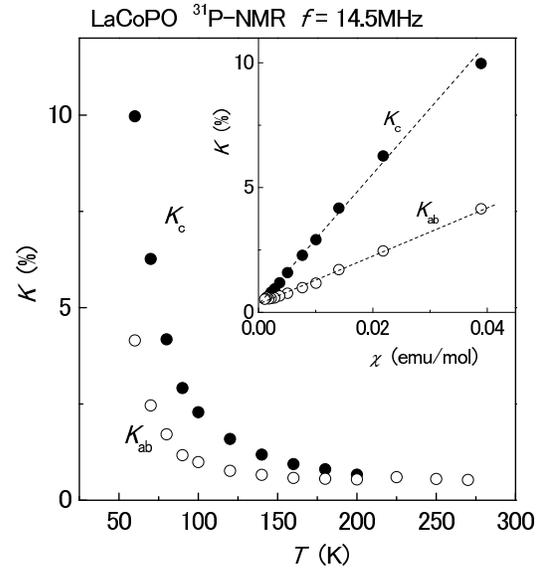}
\end{center}
\caption{Temperature dependences of the Knight shifts parallel to the $c$-axis $K_{c}$ and perpendicular to the $c$-axis $K_{ab}$. The inset shows the plots of $K_{c}$ and $K_{ab}$ against the bulk susceptibility $\chi$ measured at 0.8 T.}
\end{figure}
Figure 3 shows the temperature dependences of $^{31}K_{c}$ and $^{31}K_{ab}$, which increase as $T_{\rm Curie}$ is approached. These temperature dependences are plotted against the bulk susceptibility measured at 0.8 T, as shown in the inset. From the linear relation, the hyperfine coupling constants at the P site ($^{31}A$) are evaluated as $^{31}A_{c} = 1.55 \pm 0.02$ T/$\mu_B$ and $^{31}A_{ab} = 0.55 \pm 0.02$ T/$\mu_B$. Using these values, $^{31}A_{\rm iso} = \frac{A_{c}+2A_{ab}}{3}$ = 0.88 T/$\mu_B$ and $^{31}A_{\rm aniso}=\frac{A_{c}-A_{ab}}{3}$ = 0.33 T/$\mu_B$ are obtained, resulting in the ratio $^{31}A_{\rm aniso}/^{31}A_{\rm iso}$ = 0.37. The anisotropic term mainly originates from the 2$p$-orbital effect at the $^{31}$P site, suggestive of the presence of the coupling between Co atoms by way of the P-2$p$ orbitals. Since $A_{\rm aniso}/A_{\rm iso}$ has been reported to be $\sim 0$ in the superconducting (La$_{0.87}$Ca$_{0.13}$)FePO\cite{NakaiPRL08}, and to be $\sim 0.1$ in the stripe-type antiferromagnets BaFe$_2$As$_2$\cite{KitagawaJPSJ08} and SrFe$_2$As$_2$\cite{KitagawaJPSJ09}, the coupling through $p$ orbitals at a ligand pnictogen site in LaCoPO is stronger than that in the above-mentioned compounds, and would be important for the ferromagnetic properties.
By considering the values of $A_{c}$, the magnitude of the ordered moments at 5 K is evaluated to be $\sim 0.41~\mu_B$ from the relation $H_{\rm int}=A_{c}M_{\rm S}$, where $H_{\rm int}$ is the internal field at the P site at 5 K shown in the inset (B) of Fig.~2.  The magnitude of $M_{\rm S}$ is consistent with the magnetization result.         

\begin{figure}[htbp]
\begin{center}
\includegraphics[clip=,width=0.8\columnwidth]{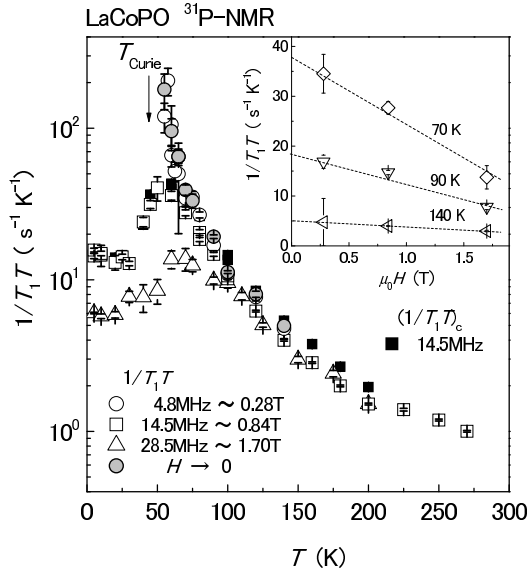}
\end{center}
\caption{Temperature dependence of $1/T_1T$ measured at the intense peak of the P-NMR spectrum at various fields. The $1/T_1T$ measured at the $c$-axis resonance field is also shown. The inset shows the field dependence of $1/T_1T$ at 70, 90, and 140 K. The values of $1/T_1T$ extrapolated to $H \rightarrow 0$ are plotted using gray circles in the main figure.}     
\end{figure}
Figure 4 shows the temperature dependence of the $1/T_1T$ measured at the intense peak in the NMR spectrum obtained at 4.8 MHz ($\sim$ 0.28 T), 14.5 MHz ($\sim$0.84 T), and 29.3 MHz ($\sim1.7$ T), which is 1/$T_1T$ along the $a$- or $b$-axis. The 1/$T_1T$ of 4.8 MHz increases on cooling and diverges at around $T_{\rm Curie}$. The divergence changes a maximum in $\sim 0.84$ T, and the maximum of 1/$T_1T$ is suppressed and the temperature of the maximum increases with increasing field, as observed in Fig.~4. This field-dependent behavior of $1/T_1T$ observed in itinerant ferromagnets, such as ZrZn$_2$\cite{KontaniJPSJ77}, Y(Co$_{1-x}$Al$_x$)$_2$\cite{YoshimuraJPSJ87}, and Sc$_3$In\cite{HiokiJPSJ77} contrasts that observed in itinerant antiferromagnets, such as LaFeAsO\cite{NakaiJPSJ08}, in which the divergent behavior of $1/T_1T$ at a N\`eel temperature is robust against applied fields. 
The $1/T_1$ measured at the $c$-axis resonance field (1/$T_1T$)$_{c}$ is plotted using closed squares in Fig.~4. It was found that $(1/T_1T)_{c}/(1/T_1T)_{ab}$ is $1.3 \pm 0.05$, indicating that the anisotropy of $1/T_1$ is small and that the relaxation rate is determined by the P-2$s$ orbitals. Since the FM critical fluctuations are easily suppressed by magnetic fields, $1/T_1T$ above $T_{\rm Curie}$ was plotted against $H$ and extrapolated to $H \rightarrow 0$ in order to determine the zero-field $1/T_1T$ related to the intrinsic FM critical fluctuations, as shown in the inset.  The $1/T_1T$ extrapolated to $H\rightarrow0$ is denoted by gray circles in Fig.~4.

\begin{figure}[htbp]
\begin{center}
\includegraphics[clip=,width=0.8\columnwidth]{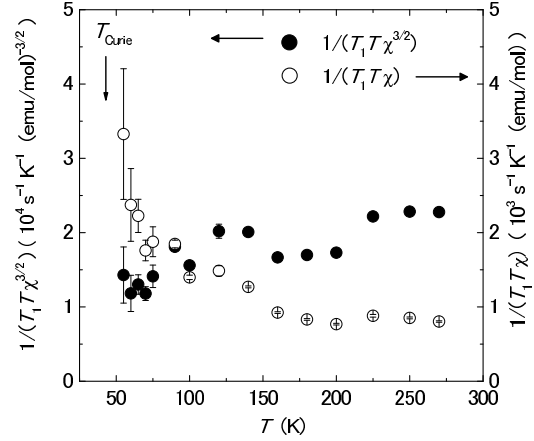}
\end{center}
\caption{Plots of $1/(T_1T\chi^{3/2})$ and $1/(T_1T\chi)$ against temperature. Here, the $1/T_1T$ extrapolated to $H\rightarrow 0$ is adopted below 150 K, and $1/T_1T$ at 14.5 MHz is above 150 K. $1/(T_1T\chi^{3/2})$ [$1/(T_1T\chi)$] is expected to be constant when ferromagnetic fluctuations possess a 2D [3D] characteristics.}
\end{figure}
Physical properties of the itinerant magnetism have been discussed on the basis of spin-fluctuation theories\cite{Moriya, TakahashiJPhys97}. These theories suggest that itinerant ferromagnets show  $1/T_1T \propto \chi^{n}$ near the FM instability with $n = 1$ in 3D and $n = 3/2$ in 2D. Figure 5 shows the temperature dependences of $1/(T_1T\chi^{3/2})$ and $1/T_1T\chi$ above $T_{\rm Curie}$, where the $1/T_1T$ extrapolated to $H \rightarrow 0$ is used below 150 K and $1/T_1T$ at 14.5 MHz is adopted since the field dependence of $1/T_1T$ is negligibly small at high temperatures. As shown in Fig. 5, the $1/T_1T \propto \chi^{3/2}$ scaling holds in the wide temperature region above 100 K, indicating that the FM fluctuations in LaCoPO possess a 2D characteristic. The gradual decrease in $1/(T_1T\chi^{3/2})$ below 100 K might be ascribed to the crudeness of the estimation of $1/T_1T$ in $H\rightarrow0$ and/or interlayer couplings inducing 3D magnetic ordering. 

\begin{figure}[htbp]
\begin{center}
\includegraphics[clip=,width=0.8\columnwidth]{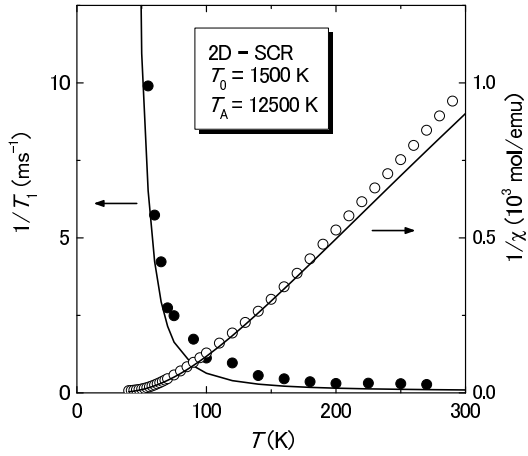}
\end{center}
\caption{$1/T_1$ extrapolated to $H \rightarrow 0$ and $1/\chi$. Solid lines are calculations based on the 2D ACRE theory. For the fitting, experimental results of $A_{\rm iso} = 1.9 $T$/g\mu_B$ and $M_{\rm S} = 0.4 \mu_B$ are adopted for $A_{\rm hf}$ and $p$, and $T_0$ = 1500 K and $T_A$ = 12500 K are evaluated.}
\end{figure}
Hatatani and Moriya extended the self-consistent renormalization (SCR) theory of spin fluctuations to 2D ferromagnets\cite{HatataniJPSJ95} and discussed the effect of spin fluctuations on various physical quantities. Here, we compare the experimental results of $\chi$ and $1/T_1T$ with their calculations and evaluate spin fluctuation parameters.
In the SCR expression, the dynamical spin susceptibility $\chi(q, \omega)$ is described in terms of two parameters, $T_0$ and $T_A$, which characterize the width of the spin excitation spectrum in frequency $\omega$ and $q$ spaces, respectively. For FM correlations as in LaCoPO, $\chi(q,\omega)$ is expressed as
\begin{equation*}
\chi(q,\omega) = \frac{\pi~T_0}{T_A}\left[\frac{x}{2\pi~T_0~x(y+x^2)-i\omega}\right],
\end{equation*}
where $x=q/q_B$ with $q_B$ being an effective zone boundary vector, and the reduced inverse susceptibility $y=1/(2T_A\chi(0,0))$.
The relation between the measured susceptibility $\chi$ (emu/mol) and $\chi(0,0)$ is $\chi=(g\mu_B)^2\chi(0,0)N_A$. 
$1/T_1T$ is related to the imaginary part of the dynamical susceptibility as
\begin{equation*}
\frac{1}{T_1} = \gamma_n^2~T\lim_{\omega \rightarrow 0} \sum_{q} [A(q)]^2 \frac{\chi''(q,\omega)}{\omega},
\end{equation*}
where $\gamma_n$ and $A(q)$ are the nuclear gyromagnetic ratio and the $q$-dependent hyperfine coupling constant, respectively. Assuming $A(q)$ is $q$-independent $A_{\rm hf}$, $1/T_1$ is calculated following the 2D formalism as 
\begin{equation*}
\frac{1}{T_1} = \frac{\gamma_n^2~A_{\rm hf}^2~T}{4T_AT_0y^{3/2}}.
\end{equation*}
In the framework of the 2D-SCR theory, $y$ is approximately expressed as $y=(T/6T_0)^{2/3}\exp{[-p^2T_A/10T]}$ for the ferromagnet ground state, where $p$ is the FM moment in $\mu_B$ units\cite{HatataniJPSJ95}. Taking the isotropic component of hyperfine coupling constant $A_{\rm iso} = 1.9 $T$ / g\mu_B$ as $A_{\rm hf}$ and $p = 0.4 \mu_B$, we calculate the temperature dependences of $1/\chi$ and $1/T_1$, as shown in Fig.~6. The experimental temperature dependences are consistently reproduced using $T_0 = 1500$ K and $T_A = 12500$ K. These values are comparable to those in Y(Co$_{0.87}$Al$_{0.13}$)\cite{YoshimuraJPSJ87} and approximately two times larger than those in LaCoAsO\cite{OhtaPRB09}. However the difference between parameters in LaCoPO and those in LaCoAsO may reflect a different model; the 3D-SCR model was used for the analyses of $\chi$ in LaCoAsO. Note that the $1/\chi$ values of LaCoPO and LaCoAsO show a downward behavior below 100 K. Since the temperature dependence of $1/\chi$ in the entire temperature range is not fitted by the 3D SCR but is reasonably fitted by the 2D SCR theory, the temperature dependence suggests the predominance of the 2D ferromagnetic fluctuations, as indicated by the relation $1/T_1T \propto \chi^{3/2}$.  However, since pure 2D ferromagnets do not have a finite critical temperature, interlayer coupling would work in the narrow temperature region near $T_{\rm Curie}$, as pointed out by Ohta and Yoshimura\cite{OhtaPRB09}. Neutron scattering experiments are required to obtain the spin fluctuation spectra along various $q$ vectors and for comparison with our NMR results.    

In conclusion, the FM critical fluctuations, which highly depend on applied magnetic fields, are observed in LaCoPO. The relation $1/T_1T \propto \chi^{3/2}$ in the zero-field limit suggests that the FM correlations possess a 2D characteristics. The temperature dependences of $\chi$ and $1/T_1T$ above $T_{\rm Curie}$ are consistently interpreted in terms of the 2D SCR theory and the spin fluctuation parameters are derived. The present data show that LaCoPO is a unique itinerant ferromagnet, in which the 2D FM fluctuations anticipated from the crystal structure are dominant in a wide temperature region above $T_{\rm Curie}$. 

We thank Y.~Maeno, S.~Yonezawa, and H.~Takatsu for experimental support and valuable discussions. We also thank H.~Ikeda, H.~Ohta and K.~Yoshimura for valuable discussions and comments.
The authors were supported by a Grant-in-Aid for Scientific Research on Innovative Areas``Heavy Electrons" (No. 20102006) from the Ministry of Education, Culture, Sports, Science and Technology (MEXT) of Japan, a Grant-in-Aid for the Global COE Program ``The Next Generation of Physics, Spun from Universality and Emergence'' from the MEXT of Japan, and Grants-in-Aid for Scientific Research from the Japan Society for Promotion of Science (JSPS). 

\bibliography{./15969.bib}

\end{document}